\documentclass[a4paper]{article}
\usepackage{amsmath}
\usepackage{amsfonts}
\usepackage{enumerate}
\usepackage{cite}
\usepackage{multicol}
\usepackage{graphicx}
\normalbaselines
\begin{document}
\title{ Higher Order Supersymmetric Truncated Oscillators }
\author{DJ Fern\'andez C$^1$ and VS Morales-Salgado$^2$\\
Departamento de F\'{\i}sica, Cinvestav,  A.P. 14-740, \sl 07000 Mexico City, M\'exico\\
{\small$^1$david@fis.cinvestav.mx, $^2$vmorales@fis.cinvestav.mx}}
\maketitle
\begin{abstract}
 We study the supersymmetric partners of the harmonic oscillator with 
 an infinite potential barrier at the origin and obtain the conditions under 
 which it is possible to add levels to the energy spectrum of these systems. 
 It is found that instead of the usual rule for non-singular potentials, 
 where the order of the transformation corresponds to 
 the maximum number of levels which can be added,
 now it is the integer part of half the order of the transformation 
 which gives the maximum number of levels to be created.
\end{abstract}
\section{Introduction}
 The so called spectral design in quantum mechanics basically 
 consists in producing a Hamiltonian with a prescribed spectrum,
 departing from an initial one whose spectrum is already known.
 Among the various techniques available to implement the spectral design,
 supersymmetric quantum mechanics has proven to be a powerful one,
 since its defining equation gives the relation between the energy spectrum 
 of the initial and new Hamiltonians (usually called supersymmetric partners) 
 in a straightforward manner                              \cite{wi81,wi82,mi84,ar85,vs93,ad94,ekk94,cks95,aicd95,bs97,srk97,fhm98,jr98,mnn98,qv99,
 sa99,ap99,acin00,b01,crf01,ast01,mr04,ac04,fe10,fgn11,qe11,ai12,ma12a,ma12b,ggm13,bsps14}.
 While first investigations focused on non-singular potentials, 
 it is important to study the way that supersymmetric quantum mechanics works for
 potentials with singular terms, e.g. infinite walls, centrifugal barriers, etc.
 \cite{ar85,aicd95,mnn98,ap99,fgn11,ma12b}.
 In this paper we will see that the so-called truncated oscillator plays 
 a special role for this kind of studies.
 
 Let us note that a truncation for the harmonic oscillator was introduced in \cite{b67},
 where the infinite-dimensional matrices representing the position and momentum operators
 were replaced by the corresponding finite-dimensional matrices generated from the first Fock states. Later, in \cite{bbkr97,b07} its parasupersymmetric partners were studied 
 and its quasicoherent states were obtained, respectively.   
 On the other hand, a different truncation, in the domain of definition of the potential,
 is also possible:
 in \cite{mnn98} it was shown that an infinite potential barrier added to 
 the harmonic oscillator modifies the domain of its supersymmetric partners,
 according with the position where the infinite barrier is placed.
 In \cite{fm14,fm15,fm16} we started to study the possibilities of spectral design 
 for the harmonic oscillator with an infinite potential barrier at the origin,
 or {\it truncated oscillator} for short.
 It was found that a first order supersymmetric 
 transformation produces only isospectral partners, 
 while a second order transformation allows to add at 
 most one level to the spectrum of the new Hamiltonian.
 Let us recall that a non-singular higher order supersymmetric transformation can be 
 decomposed as an iteration of first and second order non-singular transformations 
 \cite{as07,s08,s10}.
 This suggests that, in order to add $n$ new levels to the spectrum of the truncated oscillator,
 it is necessary to use a supersymmetric transformation of order $2n$ at least.
 
 In this work we continue the study of spectral design for 
 the supersymmetric partners of the truncated oscillator,
 by generalizing the conditions under which such singular systems
 can acquire additional energy levels in the case of a transformation of arbitrary order.
 However, instead of iterating lower order transformations we will use a straightforward 
 method that employs a single transformation of high order,
 and then we will analyze the fulfillment of the boundary conditions.
 
 This article is organized as follows:
 In Section 2 we introduce the supersymmetric transformation,
 relying on the truncated oscillator to exemplify the technique.
 In Section 3 and 4 we recall the results found previously 
 for the first and second order cases, respectively.
 In Section 5 we present the novel results of this work,
 regarding the higher order supersymmetric partners 
 of the truncated oscillator and how it turns out that 
 only a subset of the possible eigenfunctions
 for these systems actually satisfy the boundary conditions.
 Finally, in section 6 we present our concluding remarks.
\section{Supersymmetric partners of the truncated oscillator}
 A supersymmetry transformation in quantum mechanics relates two 
 Schr\"odinger Hamiltonians $H$ and $\tilde H$ through the intertwining equation
 \begin{equation}\label{intert}
  \tilde HQ=QH ,
 \end{equation}
 where $Q$ is a non-singular $k$-th order differential operator known as 
 {\it intertwining operator}.
 Once the intertwining relation has been established,
 $H$ and $\tilde H$ are called supersymmetric partners of each other.
 
 From (\ref{intert}) it follows that if $\psi_n$ is a solution of 
 the stationary Schr\"odinger equation $H\psi_n=E_n\psi_n$, 
 associated to the value $E_n$, then 
 \begin{equation}\label{neweigen}
  \tilde \psi_n\propto Q\,\psi_n
 \end{equation}
 is a solution of $\tilde H \tilde \psi_n=E_n \tilde \psi_n$,
 associated to the same value $E_n$. 
 If $\psi_n$ and $\tilde \psi_n$ both satisfy the corresponding boundary conditions,
 then we can say that there are eigenvalues $E_n$ in the spectrum of $H$ 
 which are also in the spectrum of $\tilde H$ and 
 that their corresponding eigenfunctions are related by equation (\ref{neweigen}).
 However, the functions $\tilde\psi_n$ in general do not form a complete set,
 since it is possible to find functions which are orthogonal to them \cite{mi84}.
  
 Indeed, suppose that $\phi$ is one of such functions, i.e.,
 \begin{equation}
  \langle\phi|\tilde\psi_n\rangle=0
  \quad\implies\quad
  \langle\phi|Q\psi_n\rangle=\langle Q^\dagger\phi|\psi_n\rangle=0 .
 \end{equation}
 
 Since the  set of eigenfunctions of $H$ is complete, 
 it turns out that the only vector which is orthogonal to all $|\psi_n\rangle$ is null.
 From this, we can see that the only {\it missing eigenfunctions} 
 are those contained in the kernel of the intertwining operator $Q^\dagger$.
 Since it is a $k$-th order linear differential operator,
 the dimension of such a kernel is precisely $k$.
 Even more, we can readily see that $\tilde H$ and $Q^\dagger$ commute in said kernel,
 thus they can be diagonalized simultaneously.
 
 Hence, a possible non-isospectral eigenfunction $\phi_\epsilon$ of $\tilde H$ 
 is obtained as a zero mode of the adjoint of the intertwining operator $Q$, i.e.,
 \begin{equation}\label{zeromodes}
  Q^\dagger\,\phi_\epsilon = 0 ,
 \end{equation}
 corresponding to the eigenvalue $\epsilon$.
 These are regarded as eigenvalues added to 
 the initial energy spectrum as a result of the technique.
 We can conclude that the number of non-isospectral eigenfunctions of $\tilde H$ 
 is less than or equal to the order $k$ of the intertwining operator.
 This shows that the spectral values of $\tilde H$ 
 are exhausted by two sets of possible eigenfunctions,
 the isospectral ones $\tilde\psi_n$ given by Eq. (\ref{neweigen})
 and the $k$ non-isospectral solutions satisfying Eq. (\ref{zeromodes}),
 depending on whether they satisfy the boundary conditions
 \cite{ac04,aicd95,b01,fe10,ai12}.
 
 Let us suppose now that $H=-\frac{1}{2}\frac{{\rm d}^2}{{\rm d}x^2}+V$ is the Hamiltonian 
 for the one-dimensional harmonic oscillator with an infinite potential barrier at the origin, 
 which from now on we shall call {\it truncated oscillator}, where the potential is given by
 \begin{equation}\label{potential}
  V = \left\{\begin{array}{l l}
                          \frac{x^2}{2} & \quad \text{if $\,x>0$  }\\
                          \infty  & \quad \text{if $\,x \leq 0$ }\\
          \end{array} \right. .
 \end{equation}
 
 The spectrum of $H$ consists of the set of values $E_n$ associated to
 those solutions $\psi_n$ of the stationary Schr\"odinger equation
 \begin{equation}\label{schrodinger}
  H\,\psi_n=E_n\,\psi_n\,
 \end{equation} 
 such that $\psi_n(x)$ is square-integrable in the domain $(0,\infty)$
 and it satisfies the boundary conditions, i.e. $\psi_n(0)=\psi_n(\infty)=0$.
 Therefore, the eigenvalues of $H$ are given by $E_n=2n+3/2$, $n\in\mathbb{N}$,
 with corresponding eigenfunctions 
 \begin{equation}\label{oddeigen} 
  \psi_n(x)=C_n\,e^{-x^2/2}\,H_{2n+1}(x)\, ,
 \end{equation}
 where $C_n=\left[\sqrt{\pi}\,2^{2n}\,(2n+1)!\right]^{-1/2}$ is a normalization constant
 and $H_m(x)$ is the $m$-th Hermite polynomial. 
 These are the odd eigenfunctions of the standard one-dimensional harmonic oscillator 
 now normalized in the domain $(0,\infty)$.
 
 The even eigenfunctions of the standard harmonic oscillator,
 normalized in $(0,\infty)$ and associated to the values 
 $\mathcal{E}_{n}=2n+\frac{1}{2}$ are given by 
 \begin{equation}\label{eveneigen}
  \chi_n(x)=B_n\,e^{-x^2/2}\,H_{2n}(x)\,,
 \end{equation}
 where $B_n=\left[\sqrt{\pi}\,2^{2n-1}\,(2n)!\right]^{-1/2}$. 
 These are solutions to the equation (\ref{schrodinger})
 but they do not satisfy the boundary condition at $x=0$.
 Thus, $\mathcal{E}_{n}$ are not elements of the spectrum of $H$.
 Nonetheless, we will see that they play an important role in the further 
 development of the supersymmetric technique for the truncated oscillator.
 
 Figure 1 contains a plot of the potential $V$ of the truncated oscillator.
 It also shows the eigenfunctions of the ground state $\psi_0(x)$ 
 and the first three excited states.  
 As is customary when depicting the potential and eigenfunctions in the same plot, 
 the $x$-axis for each eigenfunction is shifted 
 up to the energy of its corresponding eigenvalue.
 \begin{figure}[h]
  \centering
  \includegraphics[width=0.5\textwidth]{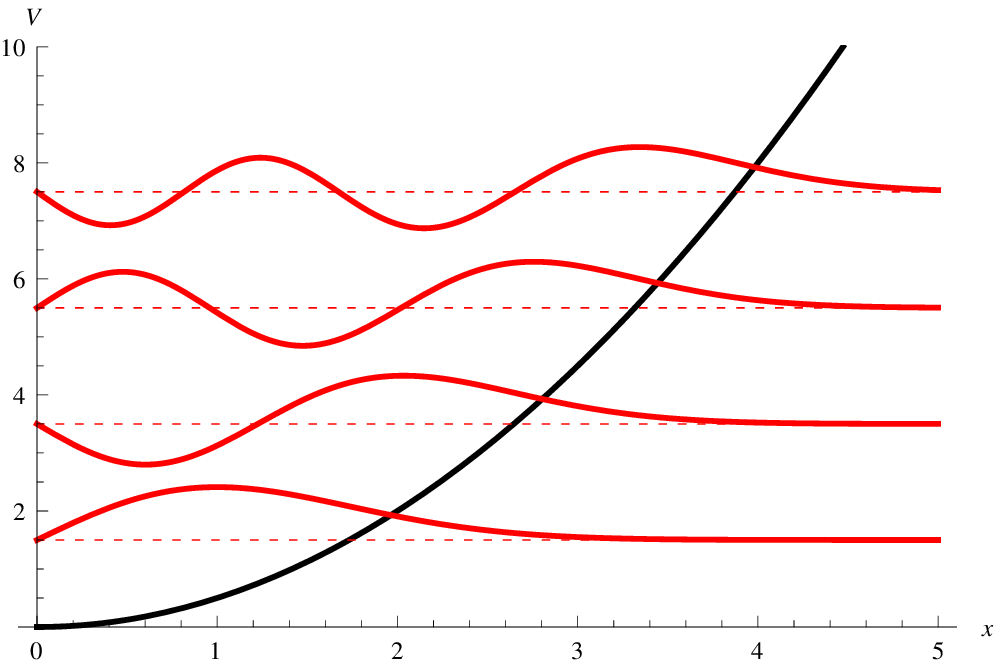}\\
  \small{Figure 1 - Potential for the truncated oscillator and 
                    the eigenfunctions of its first four energy levels.}
 \end{figure}
 \section{First-order supersymmetric partners}\label{1susy} 
 Suppose that $Q$ is a first-order differential operator  defined as
 \begin{equation}\label{supercharge}
  Q=\frac{1}{\sqrt{2}}\left[-\frac{{\rm d}}{{\rm d}x}+\alpha(x)\right] ,
 \end{equation}
 where $\alpha(x)$ is a real function of $x$ which can be obtained 
 by substituting equation (\ref{supercharge}) into equation (\ref{intert}).
 Moreover, if one assumes that $\alpha=u'/u$,
 where $u(x)$ is called the {\it seed solution}, it turns out that
 \begin{equation}\label{schrodingeru} 
  -\frac{1}{2}\frac{{\rm d}^2u}{{\rm d}x^2}+V u=\epsilon u ,
 \end{equation}
 being $\epsilon$ a parameter called {\it factorization energy},
 which we shall consider real.
 This is the stationary Schr\"odinger equation defined by $H$, 
 but $u$ is not required to be physical, i.e. to satisfy the boundary conditions.
 However, $u$ must be such that $\alpha$, and thus $Q$, is non-singular.
 If this is the case, then $V$ and $\tilde V$ can be related by
 \begin{equation}\label{newpotential}
  \tilde V = V - \frac{{\rm d}^2}{{\rm d}x^2} {\rm ln}(u).
 \end{equation}
  
 The general expression for the seed solution is a linear combination of two functions
 with opposite parity given by
 \begin{equation}\label{seedsolution}
  u(x)=e^{-x^2/2} \left[ b_1\,_1F_1\left(\frac{1-2\epsilon}{4},\frac{1}{2},x^2\right)+b_2\,x\,_1 F_1\left(\frac{3-2\epsilon}{4},\frac{3}{2},x^2\right)\right],
 \end{equation}
 where $b_1$ and $b_2$ are real constants and 
 $_1F_1(a,b;x)$ is the confluent hypergeometric function. 
 A simple way to fulfill the appropriate boundary conditions for the eigenfunctions
 of the new Hamiltonian is to proceed by cases, choosing a definite parity of $u$.
 \subsection{Odd seed solution}
 By choosing $b_1=0$ and $b_2=1$ we obtain the odd seed solution
 \begin{equation}\label{oddseed}
  u(x)=xe^{-x^2/2}\,_1F_1\bigg(\frac{3-2\epsilon}{4},\frac{3}{2},x^2\bigg)\,,
 \end{equation}
 and from equation (\ref{newpotential}) we obtain in particular that
 \begin{equation}\label{1.8} 
  \tilde V=V+1+\frac{1}{x^2}
           -\left\{\ln\left[_1F_1\bigg(\frac{3-2\epsilon}{4},\frac{3}{2},x^2\bigg)\right]\right\}'' \,.
 \end{equation}
 We can see that the terms $V$ and $1/x^2$ have a singularity at $x=0$,
 which is allowed since it lays outside the domain $(0,\infty)$.
 On the other hand, no singularities are added inside the domain $(0,\infty)$ as long as 
 $\epsilon<\frac{3}{2}$, which defines the allowed values for the factorization energy.
 In the limit where $\epsilon=\frac{3}{2}$ this value is removed from the spectrum of $\tilde H$.
 
 Therefore $\tilde H$ is isospectral to $H$ and its eigenfunctions are obtained using
 (\ref{neweigen}) as
 \begin{equation}\label{1phi}
  \tilde\psi_n(x)=\frac{C_n\,x^2e^{-x^2/2}}{\sqrt{E_n-\epsilon}}\,\left[\frac{4}{3}\,_1F_1\Big(1-n,\frac{5}{2},x^2\Big)
             +\Big(1-\frac{2}{3}\epsilon\Big)\frac{\,_1F_1(\frac{7-2\epsilon}{4},\frac{5}{2},x^2)}{\,_1F_1(\frac{3-2\epsilon}{4},\frac{3}{2},x^2)}\,_1F_1\Big(-n,\frac{3}{2},x^2\Big)\right] . 
 \end{equation}
\subsection{Even seed solution}
 Next we choose $b_1=1$ and $b_2=0$ such that
 \begin{equation}\label{evensedd}
  u(x)=e^{-x^2/2} \,_1 F_1\left(\frac{1-2\epsilon}{4},\frac{1}{2},x^2\right)
 \end{equation} 
 leading now to the potential
 \begin{equation}\label{1.12}
  \tilde V=V+1-\Bigg\{\ln\left[_1F_1\left(\frac{1-2\epsilon}{4},\frac{1}{2},x^2\right)\right]\Bigg\}''.
 \end{equation}
 This potential has a singularity at $x=0$,
 due to the barrier in the term $V$, 
 while the allowed values for the factorization energy are such that $\epsilon\leq\frac{1}{2}$.
 
 When studying the way in which the eigenfunctions of equation (\ref{schrodinger})
 are transformed we find that $Q\,\psi_n$ fails to satisfy the boundary conditions, 
 where $\psi_n$ is given by equation (\ref{oddeigen}).
 In fact, it is $Q\,\chi_n$ which solves the stationary Schr\"odinger equation 
 for $\tilde H$ and also satisfies the boundary conditions.
 Thus, the eigenfunctions of $\tilde H$ are given by
 \begin{eqnarray}\nonumber
  \tilde\psi_n(x)=\frac{B_n\,x\,e^{-x^2/2}}{\sqrt{\mathcal{E}_n-\epsilon}}\left[4n\,_1F_1\left(1-n,\frac{3}{2},x^2\right)
            +(1-2\epsilon)\frac{\,_1F_1(\frac{5-2\epsilon}{4},\frac{3}{2},x^2)}{\,_1F_1(\frac{1-2\epsilon}{4},\frac{1}{2},x^2)}\,_1F_1\left(-n,\frac{1}{2},x^2\right)\right] .
 \end{eqnarray}
 They are associated to the eigenvalues $\mathcal{E}_n$,
 which are the elements of the spectrum of $\tilde H$.

 It is worth mentioning that in both cases of the first order intertwining, 
 $H$ and $\tilde H$ are isospectral, up to an overall shift in the energy.
\section{Second-order supersymmetric partners}\label{2susy}
 Now, let us consider a second-order intertwining operator defined as
 \begin{equation}\label{2.1} 
  Q=\frac{1}{2}\left[\frac{{\rm d}^2}{{\rm d}x^2}
                     -\eta(x)\frac{{\rm d}}{{\rm d}x}+\gamma(x)\right] ,
 \end{equation}
 where $\eta(x)$ and $\gamma(x)$ are real functions of the position $x$.
 This realization of the supersymmetric technique can be obtained 
 using a pair of seed solutions of equation (\ref{schrodingeru}),
 $u_i(x)$ associated to $\epsilon_i$, $i=1,2$, 
 in the form given by equation (\ref{seedsolution}).
 
 Once the seed solutions have been fixed, 
 the potential of $\tilde H$ becomes 
 \begin{equation}\label{newpotential2}
  \tilde V=V-\frac{{\rm d}^2}{{\rm d}x^2}\ln\left[W\left(u_1,u_2\right)\right] .
 \end{equation}
 As it was done previously,
 we shall proceed by cases according to the parity of $u_1(x)$ and $u_2(x)$.
 For definiteness, we order the factorization energies as $\epsilon_1<\epsilon_2$;
 then, there are four non-equivalent identifications of these functions 
 consistent with expressions (\ref{oddseed}) and (\ref{evensedd}).
\subsection{Odd-odd seed solutions}
 Consider the identification
 \begin{equation}
  u_1=xe^{-\frac{x^2}{2}}\,_1 F_1\left(\frac{3-2\epsilon_1}{4},\frac{3}{2};x^2\right)\,,\qquad 
  u_2=xe^{-\frac{x^2}{2}}\,_1 F_1\left(\frac{3-2\epsilon_2}{4},\frac{3}{2};x^2\right)\,. 
 \end{equation}
 Then, the new potential is given by
 \begin{equation}
  \tilde V=V+\frac{3}{x^2}+2-\left[\ln w_1(\epsilon_1,\epsilon_2;x)\right]'' ,
 \end{equation}
 where $w_1$ must not have zeros in the domain $(0,\infty)$.
 To achieve this, the factorization energies must be such that
 $\epsilon_1<\epsilon_2< E_0$ or $E_n<\epsilon_1<\epsilon_2< E_{n+1}$. 
  
 The eigenfunctions of $\tilde H$ are obtained as the action of 
 the intertwining operator on the eigenfunctions of $H$:
 \begin{equation}\label{usualphi}
  \tilde\psi_n(x)=\frac{Q\,\psi_n(x)}{\sqrt{(E_n-\epsilon_1)(E_n-\epsilon_2)}} ,
 \end{equation}
 corresponding to the eigenvalues $E_n=2n+3/2$.
 In the limit when $\epsilon_2=E_0$, $E_n=\epsilon_1$ or $\epsilon_2= E_{n+1}$,
 the energy level $E_0$, $E_n$ or $E_{n+1}$ is erased, respectively;
 however, no new levels can be added to the spectrum of $\tilde H$ in this case.
\subsection{Even-odd seed solutions}
 By choosing now
 \begin{equation}
  u_1=e^{-x^2/2}\ _1 F_1\left(\frac{1-2\epsilon_1}{4},\frac{1}{2};x^2\right) ,\qquad 
  u_2=xe^{-x^2/2}\,_1 F_1\left(\frac{3-2\epsilon_2}{4},\frac{3}{2};x^2\right) ,
 \end{equation}
 it is obtained the potential
 \begin{equation}\label{2.17}
  \tilde V=V+2-\left[\ln w_2(\epsilon_1,\epsilon_2;x)\right]''\,,
 \end{equation}
 where $w_2$ has no zeros in $(0,\infty)$ as long as 
 $\mathcal{E}_n<\epsilon_1<\epsilon_2< E_n$.
 These are the allowed values for the factorization energies 
 for this even-odd choice of seed solutions.

 The eigenfunctions of $\tilde H$ are again given by expression (\ref{usualphi}),
 and they are associated to the eigenvalues $E_n=2n+3/2$.
 In the limit when $\epsilon_2=E_n$, 
 the level $E_n$ disappears from the spectrum of $\tilde H$.
 On the other hand, the level $\epsilon_1\neq\mathcal{E}_n$ 
 is indeed an eigenvalue of $\tilde H$ associated to the eigenfunction
 \begin{equation}\label{newphieo}
  \phi_{\epsilon_1}\propto \frac{u_2}{W(u_1,u_2)} .
 \end{equation}
 It means that, through the supersymmetric technique,
 we were able to add a new energy level to the initial spectrum of the truncated oscillator.
\subsection{Odd-even seed solutions}
 Now we consider the identification
 \begin{equation}
  u_1=xe^{-x^2/2}\,_1 F_1\left(\frac{3-2\epsilon_1}{4},\frac{3}{2};x^2\right) ,\qquad 
  u_2=e^{-x^2/2}\,_1 F_1\left(\frac{1-2\epsilon_2}{4},\frac{1}{2};x^2\right) .
 \end{equation}
 From this choice we obtain that $\tilde V$ is of the form given by equation (\ref{2.17}),
 with a different function $w_3$ instead of $w_2$.
 By applying the same considerations to the function $w_3$ as in the previous case,
 we arrive at the condition that the factorization energies must be such that
 $\epsilon_1<\epsilon_2<\mathcal{E}_0$ or $E_n<\epsilon_1<\epsilon_2<\mathcal{E}_{n+1}$
 in order to avoid singularities in the domain of the problem.
 Notice that this condition is different from those in the previous cases.
 
 The spectrum of $\tilde H$ is given again by the set of values $E_n=2n+3/2$,
 which are associated to the eigenfunctions of equation (\ref{usualphi}).
 It is possible to erase now the energy level $E_n$ by choosing $\epsilon_1=E_n$,
 and a new energy level $\epsilon_2$ can be added by choosing 
 $\epsilon_2\neq \mathcal{E}_{n+1}$, associated to the eigenfunction
 \begin{equation}\label{newphioe}
  \phi_{\epsilon_2}\propto\frac{u_1}{W(u_1,u_2)} .
 \end{equation}
\subsection{Even-even seed solutions}
 Finally, let us choose the seed solutions as
 \begin{equation}
  u_1=e^{-x^2/2}\,_1 F_1\left(\frac{1-2\epsilon_1}{4},\frac{1}{2};x^2\right) ,\qquad
  u_2=e^{-x^2/2}\,_1 F_1\left(\frac{1-2\epsilon_2}{4},\frac{1}{2};x^2\right) ,
 \end{equation}
 to obtain that the new potential is given by
 \begin{equation}
  \tilde V=V+\frac{1}{x^2}+2-\left[\ln w_4(\epsilon_1,\epsilon_2;x)\right]'' .
 \end{equation}
 This potential is non-singular for $x>0$ whenever that $\epsilon_1<\epsilon_2<\mathcal{E}_0$ 
 or $\mathcal{E}_n<\epsilon_1<\epsilon_2<\mathcal{E}_{n+1}$.

 For this choice of seed solutions, the eigenfunctions of $\tilde H$ are given this time by the 
 action of the intertwining operator on the solutions $\chi_n(x)$ (see eq. (\ref{eveneigen})):
 \begin{equation}
  \tilde\psi_n(x)=\frac{Q\,\chi_n}{\sqrt{(\mathcal{E}_n-\epsilon_1)(\mathcal{E}_n-\epsilon_2)}} 
 \end{equation}
 which are associated to the eigenvalues $\mathcal{E}_n=2n+1/2$.
 By choosing $\epsilon_2=\mathcal{E}_0$, $\epsilon_1=\mathcal{E}_n$ 
 or $\epsilon_2=\mathcal{E}_{n+1}$ the energy levels $\mathcal{E}_0$, 
 $\mathcal{E}_n$ or $\mathcal{E}_{n+1}$ can be erased, respectively.
 In this case, 
 no new levels can be added to the spectrum of $\tilde H$.
 
 Opposite to what happened for the first-order supersymmetric technique,
 the second order method allows to add energy levels for 
 the supersymmetric partners of the truncated oscillator.
 This is made possible by means of a particular choice of 
 the parities for the seed solutions employed in the procedure. 
 The work done in \cite{as07,s08,s10} indicates that 
 a non-singular supersymmetry transformation of arbitrary order can be obtained 
 as iterations of first- and second-order nonsingular transformations.
 Thus, in order to add more levels to the spectrum of $\tilde H$
 one should iterate the procedures described in this section and the previous one.
 This approach would require intermediate steps whose final result, 
 however, should be achieved by a direct procedure 
 involving only the initial and final expressions.
 In the rest of this paper we will describe how 
 a non-singular supersymmetry transformation of arbitrary order $k$ can be performed, 
 to add more than one level to the spectrum of the supersymmetric partners of the truncated oscillator.
\section{$k$-th order supersymmetric partners}\label{ksusy}
 The generalization of the supersymmetric technique for an intertwining operator $Q$
 of arbitrary order $k$ leads to the following relation between the potentials
 $V$ and $\tilde V$ \cite{acin00}:
 \begin{equation}\label{newpotentialk}
  \tilde V=V-\frac{{\rm d}^2}{{\rm d}x^2}\text{ln}\left[W\left(u_1,...,u_k\right)\right] ,
 \end{equation}
 where the $u_j$, $j=1,...,k$, are $k$ seed solutions of equation (\ref{schrodingeru})
 associated to the factorization energies $\epsilon_1<\epsilon_2<...<\epsilon_k$, respectively. 
 We can see that equation (\ref{newpotentialk}) appropriately generalizes equations 
 (\ref{newpotential}) and (\ref{newpotential2}). 
 On the other hand, a generalization of equations (\ref{newphieo}) and (\ref{newphioe})
 indicates that if there are new levels added to the spectrum of $\tilde H$,
 then these become a subset of the $\epsilon_j$'s associated to solutions of the form
 \begin{equation}\label{possiblephi}
  \phi_{\epsilon_j}\propto 
    \frac{W\left(u_1,...,u_{j-1},u_{j+1},...,u_k\right)}{W\left(u_1,...,u_k\right)} 
 \end{equation}
 such that $\tilde H \phi_{\epsilon_j}=\epsilon_j \,\phi_{\epsilon_j}$.
 It is the boundary conditions of the system which prevent the $k$ solutions 
 (\ref{possiblephi}) from becoming actual eigenfunctions.
 While the condition at $x\rightarrow\infty$ is satisfied naturally,
 the boundary condition at $x=0$ needs to be studied carefully.
 In what follows we are interested in finding under which conditions a supersymmetric 
 transformation yields the maximum number of non-isospectral eigenfunctions of $\tilde H$.

 In order to produce nonsingular transformations with normalizable solutions (\ref{possiblephi}),
 we shall assume that the seed solutions possess definite parity, 
 so that the $u_j$'s are of the form (\ref{oddseed}) or (\ref{evensedd}).
 Thus, let us define the parity $P$ of an even function as $+1$ and that of an odd functions as $-1$, i.e., 
 $P(f)=+1$ if $f$ is even and $P(f)=-1$ if $f$ is odd.
 
 Now, let us split the domain of the $\epsilon_j$'s 
 in connected subsets belonging to one of two classes:
 class $\mathcal{A}$ given by intervals of the form
 $(-\infty,\mathcal{E}_0)$ or $(E_n,\mathcal{E}_{n+1})$;
 class $\mathcal{B}$ given by intervals of the form
 $(\mathcal{E}_n,E_n)$.
 Such domains are illustrated in the following diagram:
 \begin{eqnarray}\nonumber
  \vdots \qquad\qquad\qquad\quad& \vdots \\ \nonumber
  --------------- & E_2=11/2\\ \nonumber
  \mathcal{B} \qquad\qquad\qquad & \\ \nonumber
  \cdots\cdots\cdots\cdots\cdots\cdots\cdots\cdots\cdots\cdots\cdots & \mathcal{E}_2=9/2\\ \nonumber
  \mathcal{A} \qquad\qquad\qquad & \\ \nonumber
  --------------- &  E_1=7/2\\ \nonumber
  \mathcal{B} \qquad\qquad\qquad &\\ \nonumber
  \cdots\cdots\cdots\cdots\cdots\cdots\cdots\cdots\cdots\cdots\cdots & \mathcal{E}_1=5/2\\ \nonumber
  \mathcal{A} \qquad\qquad\qquad & \\  \nonumber
  --------------- & E_0=3/2\\ \nonumber
  \mathcal{B} \qquad\qquad\qquad & \\  \nonumber
  \cdots\cdots\cdots\cdots\cdots\cdots\cdots\cdots\cdots\cdots\cdots & \mathcal{E}_0=1/2\\ \nonumber
  \mathcal{A} \qquad\qquad\qquad &
 \end{eqnarray} 
 
 In what follows, we will describe how to add energy levels to the supersymmetric partners of 
 the truncated oscillator in a single interval of class $\mathcal{A}$ or $\mathcal{B}$.
 If one wishes to add energy levels to distinct intervals,
 one must consider the results obtained here in each of these domains, separately. 
 
 Since the seed solutions have definite parity, then the function $\phi_{\epsilon_j}$
 given by equation (\ref{possiblephi}) has definite parity too.
 This in turn means that the fulfillment of the boundary condition 
 at the origin will depend on the parity of $\phi_{\epsilon_j}$. 
 With this in mind, let us establish first an important result.

 Let $\prod\limits_{k\mid c}u_j$ be the product of $k$ functions $u_j$ with
 definite parity, such that to $c$ of them we have inverted their parity, e.g.,
 in a term of a Wronskian the inversion of parities comes from differentiations.
 If $c$ is even, then there is no change in the parity of the product, i.e. 
 \begin{equation}
  P\left(\prod\limits_{k\mid c} u_j\right) = P\left(\prod\limits_{k\mid 0} u_j\right)\quad\text{for $c$ even.}
 \end{equation}

 This result implies that in order to have 
 the maximum number of non-isospectral eigenfunctions 
 of $\tilde H$ for a fixed supersymmetric transformation,
 the difference between the number of even and odd 
 seed solutions $u_j$ should be either one or zero. 
 If said difference is greater than one,
 then two seed solutions with the same parity can be identified 
 giving place to an intermediate second-order transformation 
 such that their corresponding solutions $\phi_\epsilon$ are not 
 eigenfunctions of $\tilde H$, as in the previous section, 
 thus reducing the number of non-isospectral eigenfunctions of $\tilde H$.
 Such a choice of parities for the $u_j$'s can be used to produce supersymmetric partners 
 which have less than $\left[\frac{k}{2}\right]$ non-isospectral energy levels.
 In any case, the isospectral part remains the same up to an overall shift in the energy.
 
 If the factorization energies are chosen in an interval of class $\mathcal{A}$,
 the transformation is non-singular whenever 
 \begin{equation}
  P(u_k)=1,\quad P(u_{k-1})=-1,\quad P(u_{k-2})=1,\dots,
 \end{equation}
 while if the factorization energies belong to an interval of class $\mathcal{B}$
 the transformation is non-singular whenever
 \begin{equation}
  P(u_k)=-1,\quad P(u_{k-1})=1,\quad P(u_{k-2})=-1,\dots.
 \end{equation}
 Also, let us keep in mind that if the factorization energies $\epsilon_j$ 
 are chosen above the ground level of the truncated oscillator ($E_0=3/2$), 
 then the supersymmetry transformation is non-singular as long as $k$ is even \cite{bs97}, 
 while for $\epsilon_j<E_0$ there is no restriction on such an order $k$. 
 
 Now let us proceed by cases, according to the class of interval which is being considered.
\subsection{Domain of class $\mathcal{A}$}
 In an interval of class $\mathcal{A}$ we have that $W\left(u_1,...,u_k\right)$
 is the sum of products of functions with definite parity.
 Each of these terms involves a change of parity of $c=\left[\frac{k}{2}\right]$ 
 functions with respect to the parity of the product $u_1...u_k$,
 due to the action of the differential operators $\left(\frac{\rm d}{{\rm d}x}\right)^i$,
 $i=0,1,...,k-1$ onto the functions $u_j$, $j=1,...,k$, i.e.,
 \begin{equation}
  P\left(\prod\limits_{k|c}u_j\right)=(-1)^c\,P\left(\prod\limits_{k|0}u_j\right) .
 \end{equation}
 Since
 \begin{equation}
  P\left(\prod\limits_{k|0}u_j\right)=(-1)^c ,
 \end{equation}
 it turns out that 
 \begin{equation}
  P\left(\prod\limits_{k|c}u_j\right)=(-1)^{2c}=1 .
 \end{equation}
 Therefore
 \begin{equation}
  P\left(W\left(u_1,...,u_k\right)\right)=P\left(\prod\limits_{k|c}u_j\right)=1 .
 \end{equation}
 
 Now, the parity of the Wronskian of the $k-1$ functions 
 resulting from deleting the $j$-th one is required.
 In order to find it, let us notice first of all that $P(u_j)=(-1)^{k-j}$.
 Moreover, the parity of the $k$-term product $u_1...u_k$ factorizes as 
 the product of the parity of the deleted function $u_j$ times the parity
 of the $(k-1)$-term product $u_1...u_{j-1}u_{j+1}...u_k$, which implies that:
 \begin{equation}
  P(u_1...u_{j-1}u_{j+1}...u_k)=(-1)^{k-j}P(u_1...u_k)=(-1)^{c+k-j} .
 \end{equation}
 As previously, the Wronskian $W\left(u_1,...,u_{j-1},u_{j+1},...,u_k\right)$
 is a sum of terms whose parities change due to a change of parity of 
 $\left[\frac{k-1}{2}\right]$ functions in each product, i.e.,
 \begin{eqnarray}
  P(W(u_1...u_{j-1}u_{j+1}...u_k))
   &=&(-1)^{\left[\frac{k-1}{2}\right]}P(u_1...u_{j-1}u_{j+1}...u_k)\\
   &=&(-1)^{\left[\frac{k-1}{2}\right]+\left[\frac{k}{2}\right]+k-j}\\
   &=&(-1)^{2k-j-1}=(-1)^{-(j+1)} .
 \end{eqnarray}
 
 Depending on the parity of the functions $u_j$, two different cases arise:
 \begin{enumerate}[(i)]
  \item If $P(u_j)=1$ then $j=k-2m$, where $m=0,1,..., \left[\frac{k-1}{2}\right]$.
        Thus:
        \begin{equation}
	  P(W(u_1...u_{j-1}u_{j+1}...u_k))=(-1)^{2m-k-1}=(-1)^{k+1}
	   =\left\{\begin{array}{cc} 
	            +1 & \text{if $k$ is odd}\\
	            -1 & \text{if $k$ is even} .
	           \end{array}\right.
        \end{equation}
  \item If $P(u_j)=-1$ then $j=k-(2m+1)$, where $m=0,1,..., \left[\frac{k-1}{2}\right]$,
        and then one has:
        \begin{equation}
	  P(W(u_1...u_{j-1}u_{j+1}...u_k))=(-1)^{2m-k}=(-1)^{k}
	   =\left\{\begin{array}{cc} 
	            +1 & \text{if $k$ is even}\\
	            -1 & \text{if $k$ is odd} .
	           \end{array}\right.
        \end{equation}
 \end{enumerate}

 Now we can conclude that if we choose the factorization energies in an interval of class $\mathcal{A}$,
 then for an intertwining of  even order $k$ the solutions $\phi_{\epsilon_j}$ such that $u_j$ is even obey
 \begin{equation}
  P\left(\phi_{\epsilon_j}\propto \frac{W\left(u_1,...,u_{j-1},u_{j+1},...,u_k\right)}{W\left(u_1,...,u_k\right)}\right)
                            = \frac{-1}{1}=-1 ,
 \end{equation}
 i.e., they satisfy the boundary condition at $x=0$.
 On the other hand, 
 if the order $k$ is odd then the solutions $\phi_{\epsilon_j}$ such that $u_j$ is odd obey
 \begin{equation}
 P\left(\phi_{\epsilon_j}\propto \frac{W\left(u_1,...,u_{j-1},u_{j+1},...,u_k\right)}{W\left(u_1,...,u_k\right)}\right)
                           = \frac{-1}{1}=-1 ,
 \end{equation}
 i.e., they comply the boundary condition at the origin.
 The remaining cases lead to solutions such that $P(\phi_{\epsilon_j})=1/1=1$
 that, in general, do not satisfy the boundary condition at the origin,
 although it is possible to have that $\phi_{\epsilon_j}(0)=0$ with $\phi_{\epsilon_j}$ even, 
 but this would happen for special and isolated values of $\epsilon_j$.
\subsection{Domain of class $\mathcal{B}$}
 In intervals of class $\mathcal{B}$ once again it is seen that $W\left(u_1,...,u_k\right)$
 is the sum of terms which are products of functions with definite parity, 
 each term involving a change of parity of $\left[\frac{k}{2}\right]=c$ functions
 compared to the parity of the product $u_1...u_k$, i.e.,
 \begin{equation}
  P\left(\prod\limits_{k|c}u_j\right)=(-1)^c\,P\left(\prod\limits_{k|0}u_j\right) .
 \end{equation}
 Since now
 \begin{equation}
  P\left(\prod\limits_{k|0}u_j\right)=(-1)^{\left[\frac{k+1}{2}\right]} ,
 \end{equation}
 it is obtained that 
 \begin{equation}
  P\left(\prod\limits_{k|c}u_j\right)=(-1)^{\left[\frac{k+1}{2}\right]+\left[\frac{k}{2}\right]}
   =(-1)^k ,
 \end{equation}
 i.e.,
 \begin{equation}
  P\left(W\left(u_1,...,u_k\right)\right)=(-1)^k
    =\left\{\begin{array}{cc} 
	            +1 & \text{if $k$ is even}\\
	            -1 & \text{if $k$ is odd} .
	           \end{array}\right.
 \end{equation}
 
 On the other hand, since now $P(u_j)=(-1)^{k-j+1}$ one arrives at:
 \begin{equation}
  P(u_1...u_{j-1}u_{j+1}...u_k)=(-1)^{k-j+1+\left[\frac{k+1}{2}\right]} .
 \end{equation}
 Hence:
 \begin{eqnarray}
  P(W(u_1...u_{j-1}u_{j+1}...u_k))
   &=&(-1)^{\left[\frac{k-1}{2}\right]}P(u_1...u_{j-1}u_{j+1}...u_k)\\
   &=&(-1)^{2\left[\frac{k+1}{2}\right]-2+k-j}=(-1)^{k-j} .
 \end{eqnarray}

 Once again, two different cases arise which depend on the parity of the function $u_j$:
 \begin{enumerate}[(i)]
  \item If $P(u_j)=1$ then $j=k-(2m+1)$ with $m=0,...,\left[\frac{k-1}{2}\right]$
        and thus:
        \begin{equation}
         P(W(u_1...u_{j-1}u_{j+1}...u_k))=(-1)^{2m+1}=-1 .
        \end{equation}
  \item If $P(u_j)=-1$ then $j=k-2m$ with $m=0,...,\left[\frac{k-1}{2}\right]$.
        Hence:
        \begin{equation}
         P(W(u_1...u_{j-1}u_{j+1}...u_k))=(-1)^{2m}=1 .
        \end{equation}
 \end{enumerate}
 
 We can conclude that, 
 for even $k$ and removing a seed solution $u_j$ which is even leads to  
 \begin{equation}
  P(\phi_{\epsilon_j})=\frac{-1}{1}=-1 ,
 \end{equation}
 which satisfies the boundary condition at the origin.
 However, if we remove and odd seed solution $u_j$ it turns out that
 \begin{equation}
  P(\phi_{\epsilon_j})=\frac{1}{1}=1 ,
 \end{equation}
 which, in general, does not satisfy the boundary condition at $x=0$.
 
 On the other hand, for $k$ odd and removing an even seed solution $u_j$ it is obtained that 
 \begin{equation}
  P(\phi_{\epsilon_j})=\frac{-1}{-1}=1 . 
 \end{equation}
 Nonetheless, 
 the boundary condition $\lim\limits_{x \to 0^+}\phi_{\epsilon_j}(x)=0$ is satisfied.
 If we remove an odd seed solution $u_j$ it turns out that
 \begin{equation}
  P(\phi_{\epsilon_j})=\frac{1}{-1}=-1 ,
 \end{equation}
 which at first sight may look like if it obeys the boundary condition at the origin.
 However, it can be seen that the odd Wronskian in the denominator induces a singularity at 
 $x=0$ which avoids that $\phi_{\epsilon_j}$ will satisfy indeed that boundary condition.

 As we can see from these results, 
 for an intertwining transformation of order $k$ with factorization energies 
 $\epsilon_j$ inside intervals of class $\mathcal{B}$,
 the solutions given by
 \begin{equation}
  \phi_{\epsilon_j}\propto \frac{W\left(u_1,...,u_{j-1},u_{j+1},...,u_k\right)}{W\left(u_1,...,u_k\right)},
 \end{equation}
 such that $u_j$ is even satisfy the boundary condition at the origin regardless the parity of $k$.
\subsection{Illustrative examples} 
 \begin{figure}[h]
  \centering
  \includegraphics[width=0.45\textwidth]{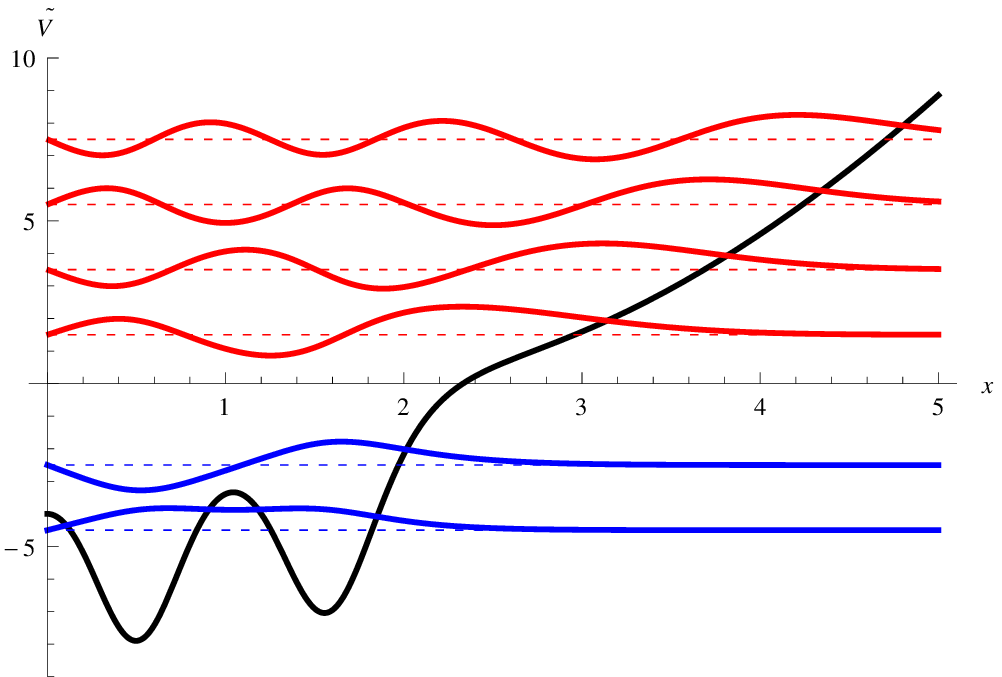}\hspace{1cm}
  \includegraphics[width=0.45\textwidth]{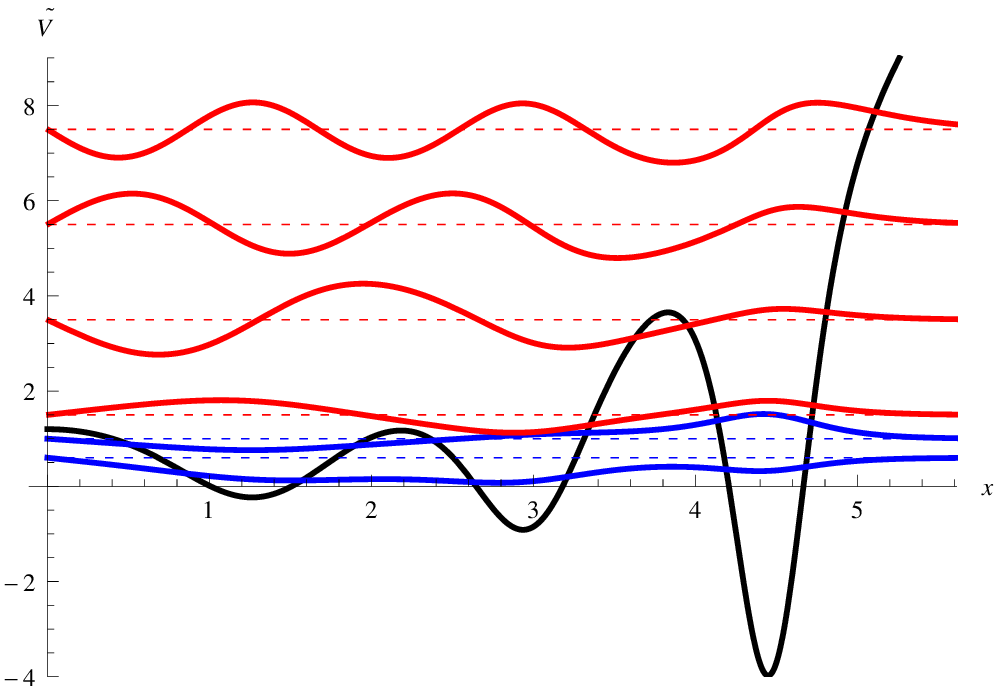}\\
  \small{(a)\hspace{8cm}(b)}\\
  \small{Figure 2 - Supersymmetric partner potentials of the truncated oscillator,
                    where two levels have been added below the ground state 
                    through a 4-th order transformation.}
 \end{figure}
 As an example of this behavior consider a fourth order 
 supersymmetric transformation with factorization energies 
 $\epsilon_1=-11/2$, $\epsilon_2=-9/2$, $\epsilon_3=-7/2$, $\epsilon_4=-5/2$,
 which belong to an interval of class $\mathcal{A}$.
 To add the maximum number of levels allowed by this transformation, 
 that is two new levels, the corresponding seed solutions are chosen such that
 $P(u_1)=-1$, $P(u_2)=1$, $P(u_3)=-1$ and $P(u_4)=1$.
 Thus, by means of equation (\ref{newpotentialk}), 
 we obtain the following supersymmetric partner potential
 {\scriptsize
 \begin{equation}
  \tilde{V}=\frac{256 x^{18}-4096 x^{16}+28416 x^{14}-99328 x^{12}+172512 x^{10}-224640 x^8+91440 x^6+86400 x^4-127575 x^2-16200}{2 \left(16 x^8-64 x^6+120x^4+45\right)^2} .
 \end{equation} }
 The two new eigenvalues added to the spectrum of the Hamiltonian $\tilde H$ 
 corresponding to this potential are $\epsilon_2=-9/2$ and $\epsilon_4=-5/2$ 
 with eigenfunctions given by
 \begin{equation}
  \phi_{\epsilon_2}=-\frac{4 \sqrt{3} e^{-\frac{x^2}{2}} x \left(8 x^6-4 x^4+10 x^2+15\right)}
                          {\sqrt[4]{\pi } \left(8 \left(2 x^4-8 x^2+15\right) x^4+45\right)} ,
 \end{equation}
 \begin{equation}
  \phi_{\epsilon_4}=-\frac{2 e^{-\frac{x^2}{2}} x \left(16 x^8+72 x^4-135\right)}
                          {\sqrt{3} \sqrt[4]{\pi } \left(8 \left(2 x^4-8 x^2+15\right) x^4+45\right)} ,
 \end{equation}
 respectively (see equation (\ref{possiblephi})).
 
 A plot of this potential and the eigenfunctions of the first six levels of $\tilde H$,
 including the two levels $-9/2$ and $-5/2$ generated below the ground state, 
 can be found in figure 2(a), depicted in a similar fashion as in figure 1.
 As for an example of a new potential with levels added in an interval of class $\mathcal{B}$,
 figure 2(b) shows a supersymmetric partner of the truncated oscillator,
 such that the factorization energies are chosen to be 
 $\epsilon_1=0.6$, $\epsilon_2=0.9$, $\epsilon_3=1$, $\epsilon_4=1.3$.
 This time the eigenvalues added to the spectrum of $\tilde H$ are 
 $\epsilon_1=0.6$ and $\epsilon_3=1$.
 In this case we omitted to show the expressions for $\tilde V$, $\phi_{\epsilon_1}$ 
 and $\phi_{\epsilon_3}$ since they result lengthy and impractical.
 However, as in the previous example,
 they can be obtained through equations (\ref{newpotentialk}) and (\ref{possiblephi}), 
 respectively.
\section{Conclusions}
 In this article we have studied the supersymmetric partners of the truncated oscillator, 
 obtained through an intertwining transformation of arbitrary order.
 As a result of this analysis we have found that in order to add new levels to 
 the energy spectrum of these systems the conditions that one must impose
 to the transformation are the following.
 
 For a supersymmetric transformation of order $k$ 
 the set of values from which we can choose the $k$ factorization energies 
 $\epsilon_1,\cdots,\epsilon_k$ can be split into two classes of intervals given by
 \begin{eqnarray}\nonumber
  \mathcal{A}&=&\left\{\left(-\infty,\frac{1}{2}\right),\left(\frac{3}{2},\frac{5}{2}\right),\left(\frac{7}{2},\frac{9}{2}\right),...,\left(\frac{3+4n}{2},\frac{5+4n}{2}\right),...\right\},\\ \nonumber
  \mathcal{B}&=&\left\{\left(\frac{1}{2},\frac{3}{2}\right),\left(\frac{5}{2},\frac{7}{2}\right),...,\left(\frac{1+4n}{2},\frac{3+4n}{2}\right),...\right\}.
 \end{eqnarray}
 Without loss of generality we shall assume that the factorization energies are ordered as
 $\epsilon_1<\cdots<\epsilon_k$ and that they are chosen in a single interval either of
 class $\mathcal{A}$ or $\mathcal{B}$.
 
 Under such assumptions we have found that in an interval of class $\mathcal{A}$ 
 the seed solutions must have parities given by $P(u_j)=(-1)^{k-j}$ while 
 in an interval of class $\mathcal{B}$ they must have parities given by $P(u_j)=(-1)^{k-j+1}$.
 Moreover, it is possible to add $\left[\frac{k}{2}\right]$ new levels 
 $\epsilon_j$ to the spectrum of $\tilde H$,
 associated to the corresponding eigenfunctions
 \begin{equation}
  \phi_{\epsilon_j}\propto \frac{W\left(u_1,...,u_{j-1},u_{j+1},...,u_k\right)}{W\left(u_1,...,u_k\right)},
 \end{equation}
 according to the following rules:
 \begin{itemize}
  \item In intervals of class $\mathcal{A}$, if $k$ is even the values $\epsilon_j$ 
        such that $u_j$ is even will be added to the spectrum of $\tilde H$,
        while if $k$ is odd and $\epsilon_1<\cdots<\epsilon_k<3/2$,
        the values $\epsilon_j$ such that $u_j$ is odd will be added to 
        the spectrum of $\tilde H$.
 
  \item In intervals of class $\mathcal{B}$, if $k$ is even the values $\epsilon_j$ 
        such that $u_j$ is even will be added to the spectrum of $\tilde H$,
        while if $k$ is odd the same values will be added to the spectrum of $\tilde H$
        whenever $\epsilon_1<\cdots<\epsilon_k<3/2$. 
 \end{itemize}

 We have seen that the truncated oscillator realizes the so called 
 spectral design of supersymmetric quantum mechanics in a peculiar manner,
 since the number of levels that can be added is bounded by $\left[\frac{k}{2}\right]$, 
 where $k$ is the order of the supersymmetry transformation. 
 Such behavior can be understood as a consequence of the singularity induced by 
 the infinite barrier which, in turn, imposes the boundary condition at the origin.
\section{Acknowledgments}
 The authors acknowledge the financial support of the Spanish MINECO 
 (project MTM2014-57129-C2-1-P) and Junta de Castilla y Le\'on (VA057U16).
 VS Morales-Salgado also acknowledges the Conacyt fellowship 243374.

\end{document}